# Chemical Diffusion at Mixed Ionic Electronic Semiconductor Interfaces and comparison with $La_2NiO_{4+\delta}$ epitaxial thin films.

J. Roqueta[1], A. Apostolidis[1], J. Chaigneau[1], R.Moreno[1], J.Zapata[1], M. Burriel[2], J. Santiso[1,3]

1: Institut Catala de Nanociencia i Nanotecnologia (ICN2) Campus UAB, 08913, Bellaterra (Barcelona), Spain

2: Department of Materials, Imperial College of London , London SW7 2AZ, UK

3: Consejo Superior de Investigaciones Científicas (CSIC) , ICN2 Building, Campus UAB, 08913, Bellaterra (Barcelona), Spain

## Abstract

A simple model to describe Mixed Ionic Electronic Conductors (MIEC) in terms of standard semiconductor physics is described. This model allows to understand defect equilibrium and charge transport at ideal heterojunctions between materials simultaneously conducting electronic and ionic point defects and to explore how rectifying effects on the electronic or ionic currents may affect the chemical diffusion and voltage at the interfaces under polarization. We found qualitatively good agreement with experimental measurements of the electrical conductivity relaxation of $La_2NiO_{4+\delta}$ thin films epitaxially grown on $NdGaO_3$ (110) substrates when the possible oxygen exchange between film and substrate is taken into account. We discuss the implications of this model to understand space charge layer formation and chemical diffusion on oxide thin film heterostructures when exposed to high temperatures and different oxygen partial pressures.

## 1. Introduction

The study of charge and mass transport properties in epitaxial heterostructures based on perovskite oxides ($ABO_3$) and layered-structure related materials is a promising field of research due to the great variety of physical phenomena that they may exhibit. The strong coupling between lattice and defect structure (oxygen sublattice) as well as possible electronic charge transfer, and modifications of spin and orbital characteristics at the interfaces may modify or give rise to novel multifunctional properties such as magnetoresistance, superconductivity, piezoelectricity or highly conducting two-dimensional electron gas at the interface as it has been demonstrated for a variety of compounds [1,2]. In some cases the origin and stabilization of these properties is still an issue of debate. Competition between different phenomena originated at the interface such as misfit strain relaxation, electronic charge transfer, or even chemical gradients (of point defects as oxygen vacancies) are claimed to influence the final properties of epitaxial thin oxide films. The role of oxygen vacancies and the distribution of other defects have been extensively studied in perovskite materials in bulk form and recently in nanosized ceramics and thin films [3]. Some of the transition metal oxide perovskites and related Ruddelsden-Popper phases exhibit high Mixed Ionic and Electronic Conductivity (MIEC) at elevated temperatures when oxygen vacancies (or interstitials) become mobile. This make these materials candidates for application as gas sensors or electrodes in Solid Oxide Fuel Cells and they have received much attention in recent years for this purpose [4]. In most of these studies it has been observed that the densities of ionic and electronic carriers near the interfaces or grain boundaries may differ from its bulk concentration due to the presence of space charged layers [5]. This makes the ionic and electronic conductivity to depend on the microstructure [6] and history of the samples and it has been currently exploited to achieve higher performance in engineered multilayers [7,8] or nanocomposites [9]. Nowadays in most of the cases the equilibrium distribution of point defects near the interfaces is well understood [10]. However, transport properties of point defects across the interfaces under non-equilibrium conditions have been investigated only for specific cases [11] although, to our knowledge, still no general theory has been developed. The main difficulty for understanding transport phenomena arises from the microscopic details of grain boundaries and interface structure that may depend from sample to sample. This situation can be avoided in highly coherent interfaces obtained from epitaxial growth where low density of charged defects and coherence in the crystal lattice is expected. Under these conditions ionic and electronic free carriers will flow from one material to the other until both species equilibrate their chemical potentials and a built-in voltage appears. The absolute value of the energies of ionic and electronic defects at both materials must be known in order to give a picture of the band alignments at the heterointerface.

In this work we provide a general formalism of Mixed Ionic and Electronic Semiconductors (MIES) as an extension of the standard semiconductor equations to include simultaneously conducting electron-holes and ionic interstitial-vacancies, which allows representing the junctions as a band diagram problem. By making use of this equivalent picture we can calculate currents across the junctions under polarization by similar analytical expressions as for semiconducting heteroepitaxial systems [12]. Special care needs to be taken in the band offset energies between materials. The development of a space charge layer at the interface seems crucial in order to reproduce some experimental features observed in the electronic conductance of oxide thin films when measured at elevated temperatures. In this work we focus on the case of $La_2NiO_{4+\delta}$ epitaxial thin films grown on $NdGaO_3$ (110) substrates (NGO) in order to compare theory and experiment.

## 2. Experimental

The $La_2NiO_4$ films were deposited either by metalorganic vapor deposition as described in a previous work [13] or by pulsed laser deposition. No substantial differences in the crystal quality of the films were detected between those two deposition techniques. Electrical conductivity relaxation (ECR) was measured for these films at high temperatures upon sequential changes of the gas atmosphere from $O_2$ to $N_2$. The conductivity was measured by depositing two parallel Ag contacts in the film surface. Transient voltages between film and a bottom Ag electrode painted in the back side of the substrate were also measured during the oxygen gas exchange. The experimental observations were modeled by numerically solving the Fick's diffusion equations described in the following paragraph and by the semiconductor model presented in sections 3.2 and 3.3. These simulations were made by ad hoc code (C+ language) using finite differences and will be described briefly on section 3.4.

## 3. Results and Discussion

### 3.1. Preliminar experimental evidences

$La_2NiO_{4+\delta}$ compound is the *n*=1 member of the Ruddelsden-Popper $La_{n+1}Ni_nO_{3n+1}$ series and presents significant electronic and ionic conductance. Under standard conditions the material may show oxygen overstoichiometry up to $\delta$=0.18 [14] located in interstitial sites with a high oxygen mobility.

This material shows also large *p*-type electronic conductivity due to the mixed Ni 2+/3+ valence state of Nickel ions. Upon changing the oxygen partial pressure at elevated temperatures, the material exchanges oxygen with the atmosphere and varies the valence state of the Nickel ions. Consequently, a change in the electronic carrier density, and therefore in the conductance, is observed. In Electric Conductivity Relaxation experiments (ECR) the conductance response is monitored after sudden changes of the $P(O_2)$. In this study we recall some ECR measurements performed on thin films of $La_2NiO_{4+\delta}$ epitaxially grown on $NdGaO_3$ substrates reported in a previous work [13]. Fig. 1a shows the sheet conductivity at 574 °C of a 350 nm thick film upon cycling from pure $O_2$ to $N_2$ gas atmospheres, and left for several hours under steady atmosphere. The conductivity decreases or increases very rapidly under $N_2$ and $O_2$ atmosphere, respectively, as expected for the fast oxygen surface exchange rate at this elevated temperature. It remains at stable values under steady $O_2$ atmosphere. However, under steady $N_2$ atmosphere the conductivity showed a continuous drift for several hours. When exposed back to oxygen atmosphere it recovered a higher conductivity state but it showed an apparent non-reversible decrease of the total conductivity of about -18%.

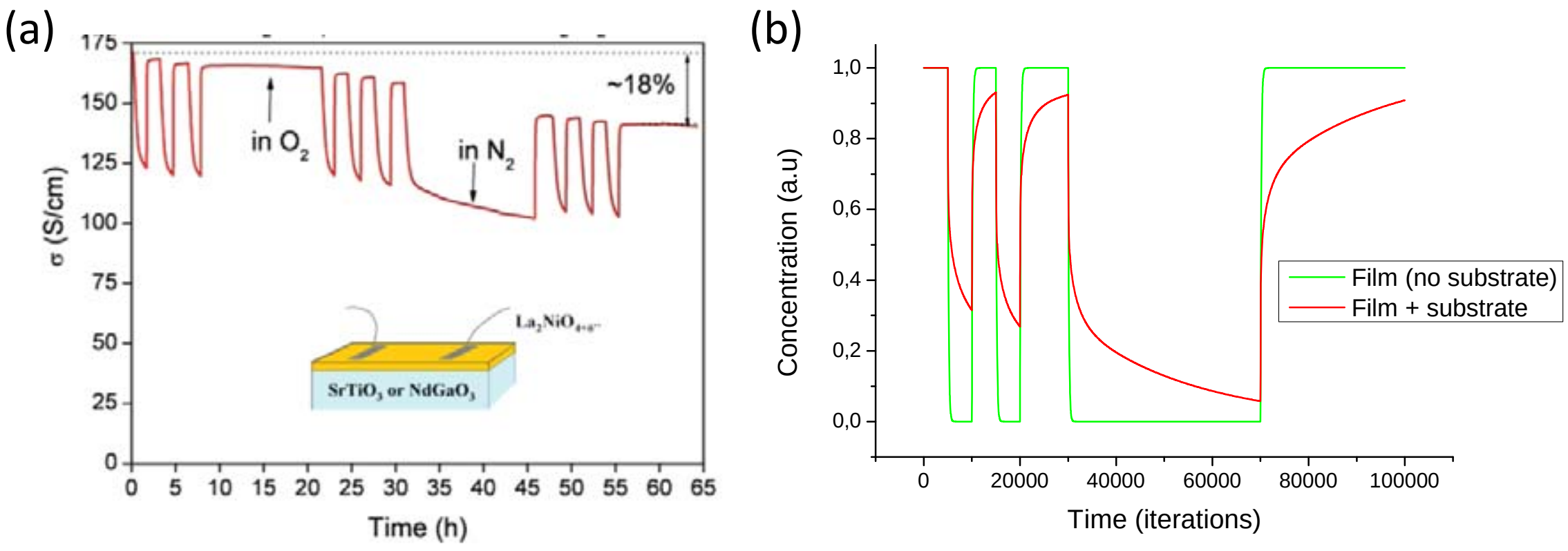

**Figure 1 : (a) Electric Conductivity Relaxation experiments on $La_2NiO_4$ thin films deposited on NGO substrates measured at 574 °C, extracted from ref.: [13]. The conductivity is measured with an AC signal of amplitude 50mV and a frequency of 1 kHz. (b) Simulation of the total concentration of defects in the film when cycling the atmosphere. In red it is shown the concentration of ionic defects in the film when exchange with the substrate is allowed. In green, concentration when no substrate is present. Simulations where performed by numerical integration of the Fick's diffusion equation.**

Since XRD analysis of the samples did not show any substantial degradation of the samples after exposure to these temperature and pressure conditions we focus on a possible interaction with the substrate to explain the overall conductivity variations. At elevated temperatures $NdGaO_3$ may present non negligible ionic conductivity (oxygen vacancies) as well as electronic conductivity (holes), as some of the insulating perovskites used as substrates (i.e. $SrTiO_3$), normally related to the presence of a certain concentration of acceptor-type impurities (i.e. Fe, Mn dopants) [8]. One may expect that the substrate material changes its oxygen stoichiometry, and therefore the concentration of electronic carriers, when the atmosphere is changed. This process can be very slow due to diffusion across the large volume of substrate material as well as the low oxygen exchange rate with the surrounding atmosphere. However, the presence of a thin film of material with a high oxygen exchange rate, such as $La_2NiO_4$, may exert a catalytic effect and facilitate the reduction of the substrate through the film, by injecting oxygen through the film-substrate interface, thus preventing the film to reach the equilibrium until the whole system has been reduced. Although the contribution of the substrate to the overall conductance may be negligible the time dependence of the film conductance could be substantially modified. Fig. 1(b) shows the results of the simulation of Fick's diffusion equation of a thin film when taking into account possible oxygen flow across the film-substrate interface. When the exchange rate between film and substrate is larger than the exchange rate with the atmosphere significant differences can be found. The apparent exchange rate (fast exponential decay) seems to be much slower and the change in conductivity (charge carrier concentration) is no longer following a simple exponential decay but needs a double exponential to be fit as a convolution with the substrate oxygen diffusion time dependence.

Still, at this point this phenomenon might be reversible and the conductivity values are expected to be fully recovered. Therefore, in order to explain the observed progressive decay on conductivity it is necessary to introduce a certain asymmetry between the oxidation and reduction processes. According to charge neutrality conditions the oxygen transport inside these materials occurs by the simultaneous flow of oxide ionic species and electrons. If one introduces a rectifying effect at the film-substrate interface for one of the charged species this will allow oxygen to flow in the forward direction (namely, from the substrate to the film) while it will prevent it to flow in the backward direction, thus allowing the substrate to be reduced across the interface but not re-oxidized. After one cycle, the substrate oxygen stoichiometry will change as well as the amount of electronic carriers near the film interface, generating a built-in voltage and producing history dependent measurements.

### 3.2. Mixed Ionic Electronic Semiconductor Model

Following the semiconductor equivalence for point defects in mixed ionic-electronic conducting materials (MIEC) already described by J. Maier [5, 10] we assume that a MIEC material can be modeled using standard semiconductor physics for both, electrons and ions simultaneously. We will treat mathematically as equivalent both the electronic and ionic carriers and we will not enter into the microscopic justification of these ideas. This study is devoted to explore the validity of this approximation. The main approximations are the following: i) there are two well defined energy levels for electrons and ions called Conduction Band and Valence Band; ii) no neutral species are mobile, (no neutral oxygen vacancies); iii) for simplicity reasons and to keep the reasoning simple we use Maxwell-Boltzmann statistics and iv) only single ionized ionic defects are considered.

The top of the valence band for electrons $E_{ec}$ is defined as the energy necessary to remove one electron from the perfect material to the vacuum creating an electronic hole defect and the bottom of the conduction band $E_{ev}$ as the energy that an electron releases when it is incorporated from the vacuum into a perfect material creating a conducting electron defect. The same treatment can be used for ions. The top of the "valence band" for ions $E_{iv}$ is the energy necessary to remove one oxygen ion from the lattice to the vacuum (leaving one oxygen vacancy), and the bottom of the "conduction band" $E_{ic}$, the energy necessary to introduce an interstitial oxygen from the vacuum. In Figure 2 it is depicted the energy levels for electrons and ions with respect to the vacuum level, the arrows indicate the fundamental processes to create defects. Ions and electrons in vacuum may react to form a neutral gas phase.

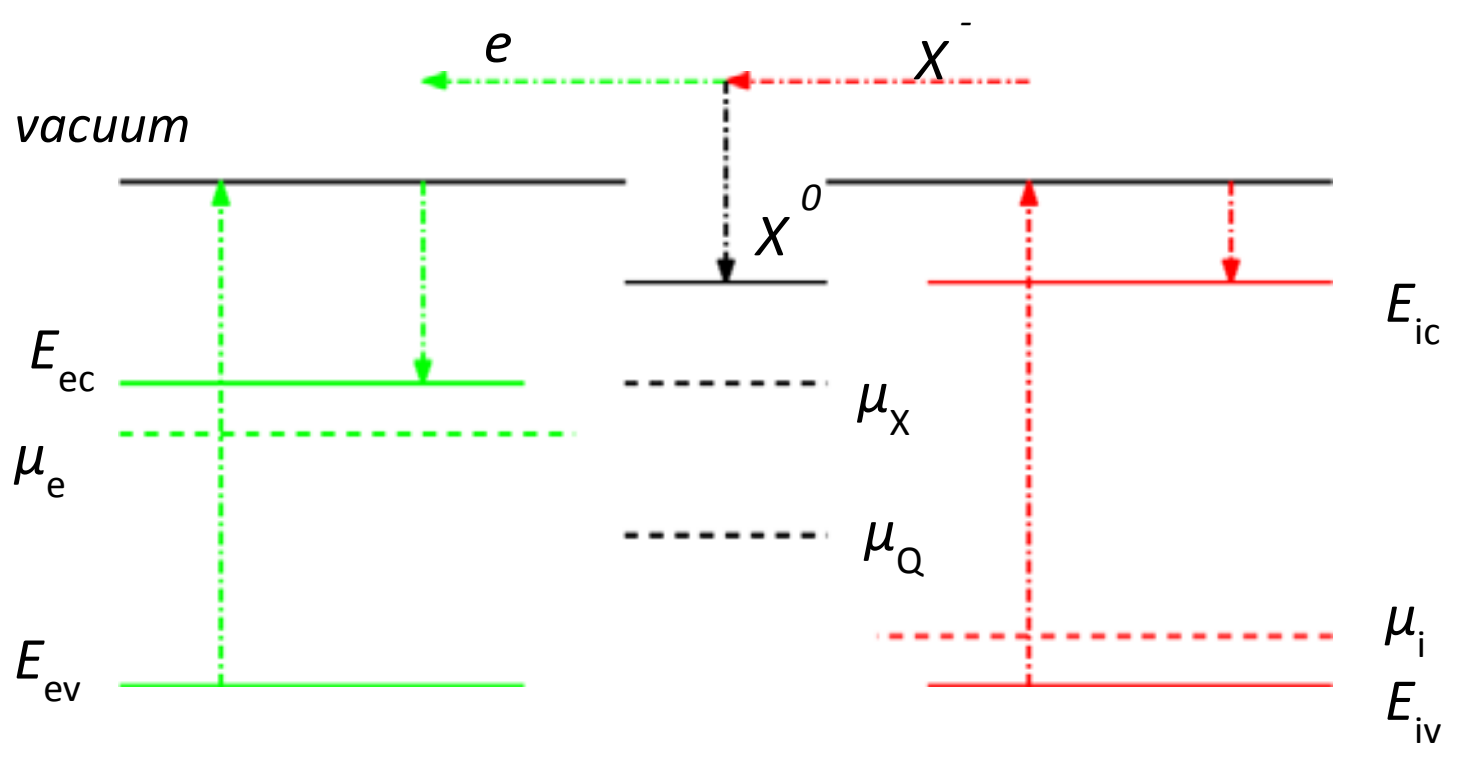

**Figure 2: Energy levels for electrons $E_{ec}$, $E_{ev}$ (green) and ions $E_{ic}$, $E_{iv}$ (red) refereed to a common vacuum level. The energy level of the neutral atomic gas phase relative to the charged ion is also shown (black). The electronic $\mu_e$ , ionic $\mu_i$ chemical potentials are depicted (green and red dashed lines, respectively) as well as the chemical potential of the gas phase $\mu_X$ and charged potential $\mu_Q$ (see text).**

$$X^{-}{}_{gas} \leftrightharpoons X^{0}{}_{gas} + e^{-}$$

The energy released in this reaction, $E_X$, can be understood as an offset between the electronic and ionic vacuum levels, although on the rest of the paper we will neglect this offset for simplicity.

The density of carriers is expressed in terms of the chemical potential for electrons and ions, $\mu_e$, $\mu_i$ as in standard semiconductors.

$$n = N_{ec}Exp(-(E_{ec} - \mu_e)/kT)$$

$$p = N_{ev}Exp(-(\mu_e - E_{ev})/kT)$$

$$I = N_{ic}Exp(-(E_{ic} - \mu_i)/kT)$$

$$V = N_{iv}Exp(-(\mu_i - E_{iv})/kT)$$

(1)

Where $n$, $p$ , $I$ , $V$ are respectively the density of electrons, holes, ion interstitials and ion vacancies, and $N_{ec}$, $N_{ev}$ , $N_{ic}$ , $N_{iv}$ are effective densities of states or simply the degeneracy of each energy level per unit volume. The logarithm of the concentration of each defect can be visualized in Fig2 as the distance between the energy level for that particular defect and the corresponding electrochemical potential.

The material must remain neutral so the electroneutrality condition must be fulfilled including the concentration of all possible ionized impurities acting as Acceptors ($A$) and Donors ($D$)

$$A - D = V + p - n - I$$

(2)

In order to express the amount of carriers as a function of the pressure of the gas phase and the doping concentration we found useful the following change on variables.

$$\mu_X = \mu_i - \mu_e$$

$$\mu_Q = \mu_i + \mu_e$$

(3)

Where the first term $\mu_X$ is the chemical potential of the gas phase and it is related to the gas pressure by using $P(X) = P_0 Exp((\mu_X - E_X)/kT)$, and can be visualized in Fig2 as the distance between ionic and electronic chemical potentials. The second term $\mu_Q$ is called the "charged chemical potential" in analogy with spintronics [15] and expresses the absolute position of chemical potentials with respect to vacuum. Note that $\mu_X$ does not contain any information of the voltage in the sample, the voltage appears only in $\mu_Q$. By inserting this expression into eqs (1) and substituting into eq (2) one obtains a relation between the amount of charged impurities and the position of these potentials. $A - D = f(\mu_X, \mu_Q)$.

Assuming a constant impurity content this equation relates $\mu_X$ and $\mu_Q$ and it is possible to obtain the equilibrium concentration for each defect at any temperature and gas pressure, or what it is known as the defect chemistry diagram.

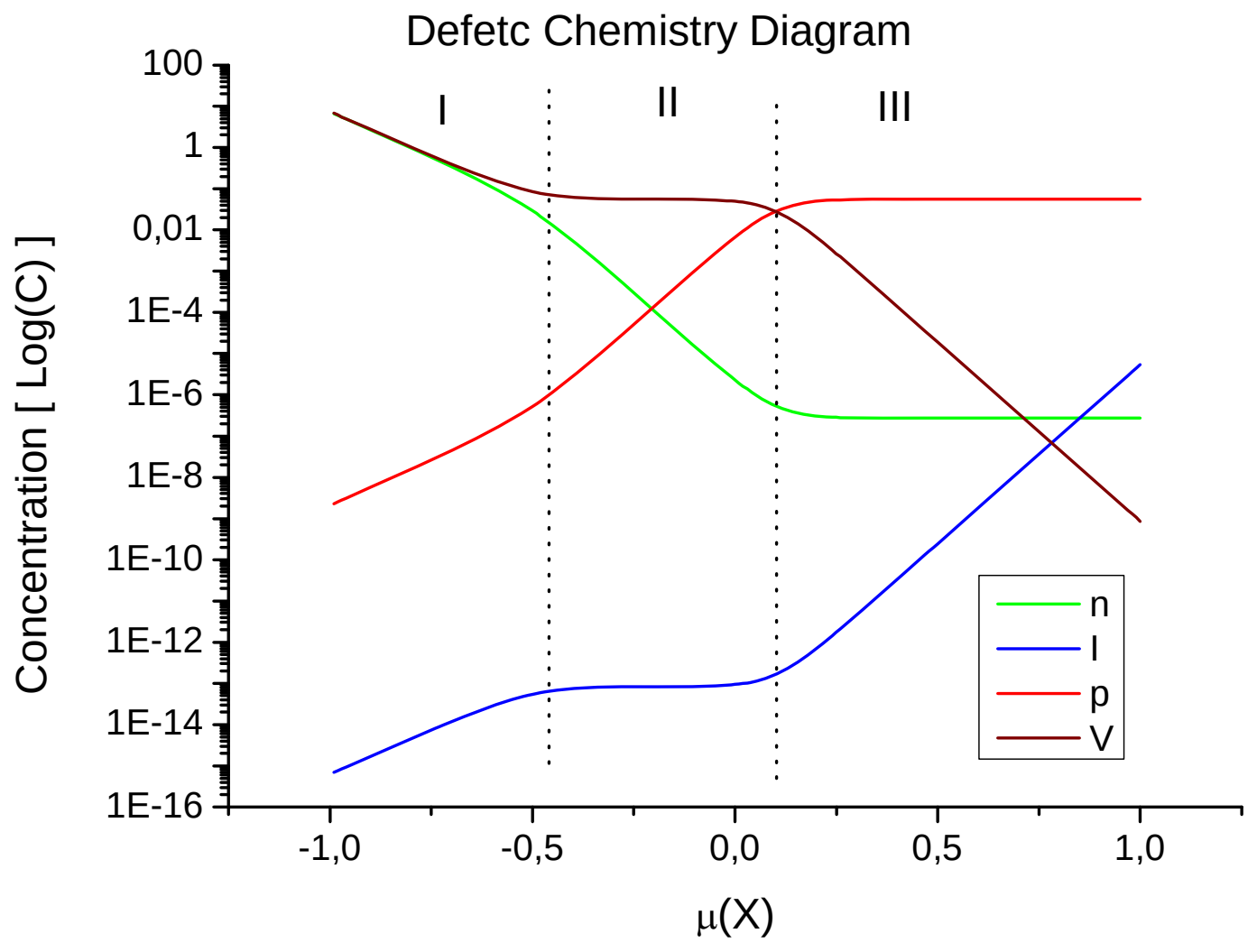

**Figure 3: Defect chemistry diagram for an Acceptor-doped material. Selected values for $E_{ec}$=1eV, $E_{ev}$=-0.4eV, $E_{ic}$=3eV, $E_{iv}$=-0.2eV, kT=0.1eV, The amount of ionized impurities is calculated at $\mu_X = \mu_Q = 0$ and maintained constant.**

We can reproduce simple defect diagrams like in [16]. Our approach is slightly different than in [16] as it introduces absolute energy levels instead of the mass-action laws, this will enable us to compare two different materials. Using eq. (1) to relate the concentration with the chemical potential may help to simplify some calculations or to introduce a particular dependence of the charge of impurities with oxygen pressure.

In Fig. 3 we plot the defect chemistry diagram of an acceptor-doped material (i.e.: typically in perovskite type substrates like $SrTiO_3$ and $NdGaO_3$, this type of impurities are the dominant ones). Three different regimes may be distinguished. At very low pressures of the gas phase, in region I, the main defects are ion vacancies and electrons in the same concentration ($n \sim V$) whereas holes and interstitial ions are kept at minority levels. We may call the material in this situation an Intrinsic MIES as the impurities have no effect on the concentration of carriers. Upon gas pressure increase, in region II, the impurities become neutralized by ion vacancies ($V \sim A$). We refer to the material in this situation an acceptor-doped MIES with oxygen vacancies as the majority carrier species. In this region we refer to electrons or holes as secondary carriers and their concentrations depend on the oxygen pressure. In region III, at higher $P(O_2)$, the impurities become neutralized by electronic holes, which are the majority species ($p \sim A$), the secondary carriers will be the ionic vacancies. Interstitials (which are supposed to have a very large energy of formation) are considered minority carriers in all regions except eventually at very high pressures when another intrinsic regime will be found. It is important to note that the concentration of majority carriers is always fixed by the dopant concentration, so their chemical potential is unaffected by gas pressure changes (inside regime II). All the changes in the pressure will mostly affect the concentration of secondary carriers.

### 3.3. Charge Transport

Charge transport equations of MIEC materials under polarization have been studied extensively by I. Riess [17, 18]. The charge and mass transport across the junction itself has remained elusive for MIEC materials while there are several approximations for semiconductors. Here we review the transport equations in a general form. These equations are needed in order to simulate the whole system by numeric integration using the finite differences procedure.

The current intensities of each species are calculated using the Drift-Diffusion equation.

$$j_i = -\upsilon_i kT \frac{dC_i}{dx} + z_i \upsilon_i C_i E \tag{4}$$

Where $j_i$ is the current density $C_i$ the concentration, $\upsilon_i$ is the mobility and $z_i$ is the charge of the ith specie and $E$ is the electric field at any particular *x* position.

The change in concentration with time of a particular defect at a specific position is calculated using the mass conservation law. We take explicitly into account the thermal generation and recombination of complementary species, electron-holes or vacancy-interstitials.

$$\frac{dC_i}{dt} = -\nabla j_i - r_{ij}(\, C_i C_j - C_{i0} C_{j0})$$

(5)

Where $r_{ij}$ is the recombination rate between species *i* and *j* and $C_{i0}$ $C_{j0}$ are the equilibrium concentrations.

The Electric field inside the material is calculated by solving the Poisson equation. In the specific planar geometry of a thin film interface the concentration is expected to vary only in the normal direction of the interface or x direction being constant in the in-plane directions y and z. Under this geometry Poisson equation is simplified and can be integrated directly.

$$E(x) = E_0 + \frac{1}{\varepsilon} \int_0^x Q(x') dx'$$

(6)

Where $Q = V + p - n - I + D - A$ is the net charge at a given position and $\varepsilon$ the dielectric constant.

At the interface between the two materials the current cannot be described by the drift diffusion equation as the diffusion coefficient is not defined. In order to take into account the possible band offsets for each defect we express separately the contributions of the forward and backward currents.

$$j_{1\to 2} = A\, C_i^{(1)}$$

$$j_{2\to 1} = A C_i^{(2)} Exp(\,-\Delta\varphi / kT)$$

(7)

where $\Delta\varphi = E_i^{(1)} - E_i^{(2)}$ is the band offset or difference in formation energies of the *i*-th specie taken into consideration.

The forward current $j_{1\to 2}$ is the current from the material (1), with high formation energy per defect, to the material (2) with lower formation energy. In this situation the carriers can diffuse directly into the second material as they lower the energy. We can consider the forward current to be proportional to the density of carriers in side (1) $C_i^{(1)}$ and by a factor $A$ that depends on the local diffusion coefficient close to the interface. In the backward direction, $j_{2\rightarrow1}$, the carriers diffuse into a material with higher energy per defect and they must overcome a barrier by thermal excitation. The probability for jumping is then reduced by a factor $Exp(-\Delta\varphi/kT)$. This can be justified by applying detailed balance. In equilibrium these two currents must be equal in magnitude and opposite in direction. Since the concentration of species in both sides differs by a factor of $Exp(\Delta\varphi/kT)$ by construction the probability for jumps must cancel out this difference in concentration. The magnitude $j_0 = |j_{1\rightarrow2}| = |j_{2\rightarrow1}|$ is called the leakage current or thermal current and it is the current of exchange of ions or electrons in equilibrium. Any nonequilibrium injection of carriers in the material (1), in the minority side, will produce a large injection through the junction. While the same variation on the other side (2), where they are majority carriers, will produce a small variation of the current as the total concentration of carriers would be almost unaffected.

Finally, we discuss the boundary condition in contact with the atmosphere. In the surface of $La_2NiO_4$ the following reaction takes place $X_{gas} \leftrightarrows I + h$

$$j_I = j_p = K_+ P(X) - K_- I p \tag{8}$$

Where $K_+$ and $K_-$ are the forward and backward reaction rates and $j_I$ , $j_p$ are the interstitial and hole currents at the boundary. $P(X)$ is the pressure of the gas phase, $I$ and $p$ are the concentrations of the interstitials and holes. For the substrate we consider that there is no direct exchange with the atmosphere.

### 3.4. Computer simulation results

The preceding coupled differential equations with boundary conditions were integrated using the finite differences method using a C++ code. The x spatial direction is discretized in 150 pixels. The first 20 pixels correspond to the film while the rest corresponds to the substrate. The concentration of each species in each pixel is saved in an array of 150 entries. The code consists on an initialization function and a main loop that reflects the evolution of the system with time. The initialization function calculates the concentration of each species in any position once the chemical potentials and the energy levels are fixed for the film and substrate. Using electroneutrality equation it is possible to calculate the impurity content in each pixel. The main loop for the time integration consists of three different steps. First, the currents are calculated from the concentration profile and electric fields using eq. 4 and using eq. 7 and eq. 8 at the interfaces. Then the concentration is recalculated using eq. 5. Finally the electric field is calculated using eq.6 . At some time intervals (after thousands of iterations) the gas pressure is changed at the boundary condition using eq. 8.

The values of each energy level were adjusted to present a defect chemistry diagram qualitatively equivalent to the $La_2NiO_4$ (intrinsic with interstitial oxygen and holes) and $NdGaO_3$ (acceptor-doped with vacancies as majority species). The specific values for the formation energies are not realistic. As most of the parameters of the model are unknown, they are set arbitrarily to 1, like the density of states, mobilities. Electric permeability was set to 10 just to enlarge the space charge layer. The purpose of the simulations is to study the effect of a rectification in one of the charged species (ions or electrons) on ECR experiments on thin films, as previously discussed. The rectifying effect will be induced by introducing a large band offset in the ionic valence band.

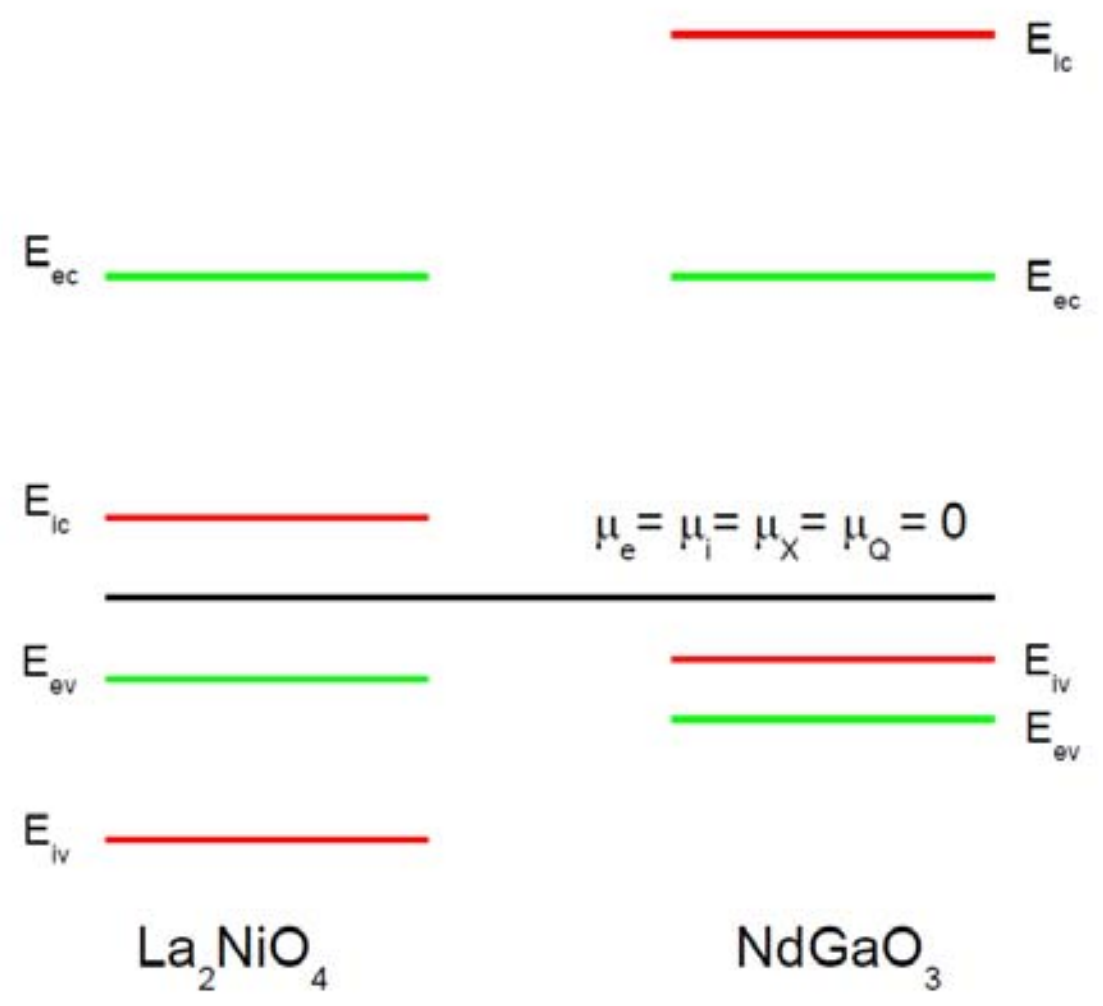

**Figure 4: Position of the electronic energy levels (Green) and ionic energy levels (Red) in each side of the junction. The energy levels are adjusted with respect to the position of chemical potentials set to $\mu_X=\mu_Q=\mu_i=\mu_e=0$ as initial conditions. On the Left proposed band diagram for the film of $La_2NiO_4$ with $E_{ec}$=1eV, $E_{ev}$=-0.3eV, $E_{ic}$=0.3eV, and $E_{iv}$=-1.3 eV, In the right the proposed band diagram for the substrate $NdGaO_3$ with $E_{ec}$=1eV, $E_{ev}$=-0.4eV, $E_{ic}$=3eV, $E_{iv}$=-0.2eV.**

To initiate the system we first fix the chemical potentials as depicted in Fig.4. We set $\mu_X$= 0 in both film and substrate, which guarantees that they are in equilibrium with the same gas phase. We also set the same charging potential $\mu_Q$ = 0, which guarantees that both materials are in equilibrium under charge transfer, i.e. "flat band" conditions. Once chemical potentials are fixed to 0, we can adjust the number of charge carriers simply by fixing the position of the valence and conduction bands for each defect. For the film we set $E_{ev}$ = -0.3eV and $E_{ic}$ = +0.3eV in order to obtain the same amount of holes (*p*) and Interstitials (*I*) and the energy for minority carriers are set to $E_{ec}$ = 1eV and $E_{iv}$ = -1.3eV. For the substrate we place the ionic valence band close to the position of the chemical potentials $E_{iv}$ =-0.2eV in order to guarantee that ion vacancies will be the majority carriers. Then we place the electronic valence band at $E_{ev}$ = -0.4eV to ensure that electronic holes are the secondary carriers. Finally, we bring the electronic and ionic conduction bands far away from the chemical potential levels $E_{ec}$ = 1eV and $E_{ic}$ = 3eV in order to keep them as minority carriers. The amount of impurities, (A-D) was set equal to the number of vacancies for the substrate and 0 for the film. The value for *kT* was set to 0.1 eV. With these initial values we ensure a flat band initial condition in equilibrium avoiding transient effects such as the formation of the space charged layer and very slow voltage-chemical couplings.

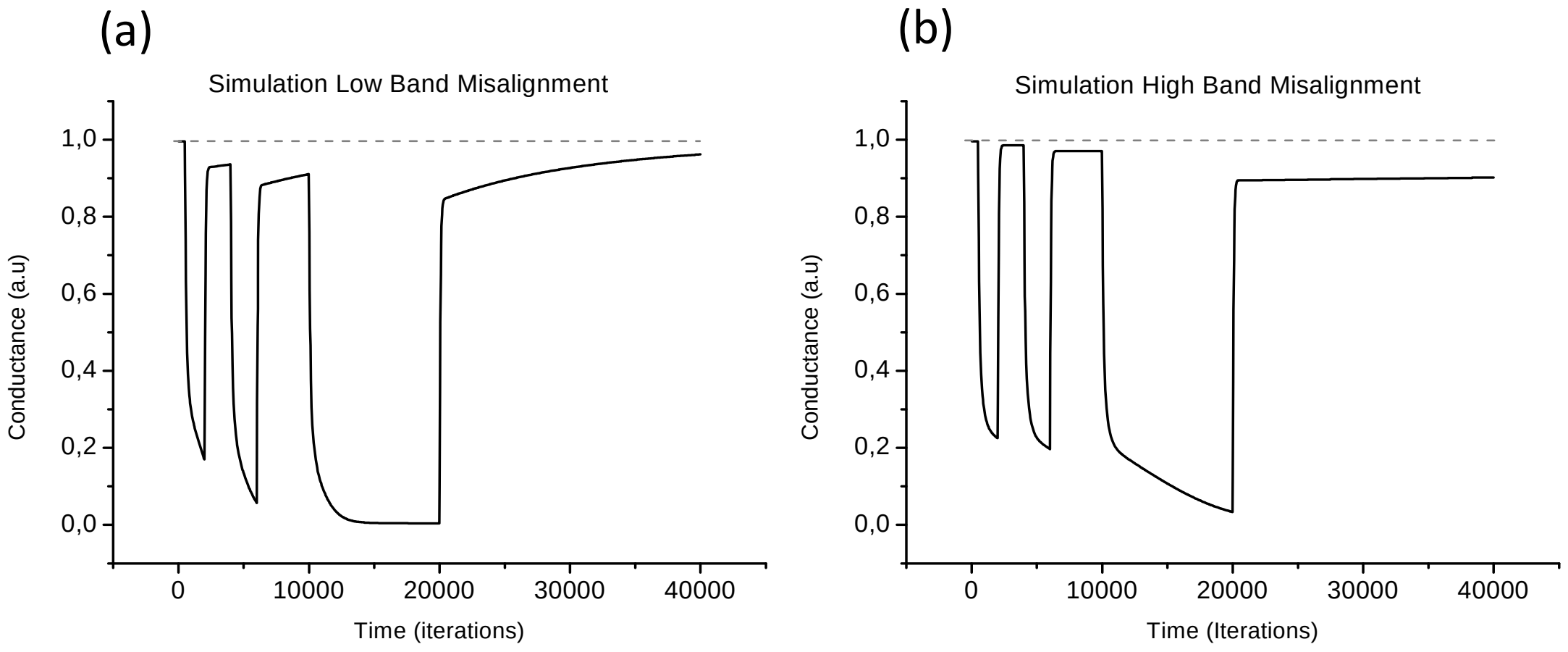

**Figure 5: Simulation of ECR experiments with different band misalignment. The energy levels for the substrate are the same as in Fig. 4 but for the film it is changed the energy for ionic valence band. In a) $E_{iv}$=-0.8 eV while in b) $E_{iv}$= -1.3 eV. The change in pressure corresponds to changes on $\mu_X$ from 0 to -0.2.**

The conductance of the film is calculated after integration of the amount of holes in the layer by assuming an arbitrary constant value of the hole mobility. We simulated the response of the system when cyclically changing the pressure from $\mu_X$= 0 to $\mu_X$= -0.2. In Fig. 5 we show the results of the simulation of conductivity relaxation experiments with two different band offsets. In (a) the energy for the ionic vacancies in the film is $E_{iv}$ = -0.8 eV and the band offset between vacancies in the film and substrate is low. In (b) this energy is reduced to $E_{iv}$ = -1.3 eV increasing the band offset.

In these simulations it is clear that relaxation of the conductivity takes place in two steps with very different response times. When the atmosphere is changed a fast response corresponding to a change on film stoichiometry is observed, but the system do not reach the equilibrium because of the interaction with the substrate. The slow response is caused by the reduction of the substrate itself that progressively alters the amount of electronic hole carriers in the film.

In the case of low band offset, in Fig. 5 (a), the system tends to reach equilibrium quite fast. In this case the amount of minority vacancies in the film is large and this allows large oxygen leakage current at the interface. When the system is exposed to oxidation conditions, it recovers the original value although a very asymmetric response is observed. When lowering the vacancy energy level, as in in Fig. 5(b) with a larger offset, the amount of minority oxygen vacancies in the film is reduced and the leakage currents across the junction are much lower. Therefore the system needs more time to equilibrate. Under oxidizing conditions the oxygen vacancies are efficiently blocked by the band offset. The substrate remains reduced for very long time and the film conductivity does not recover the original values.

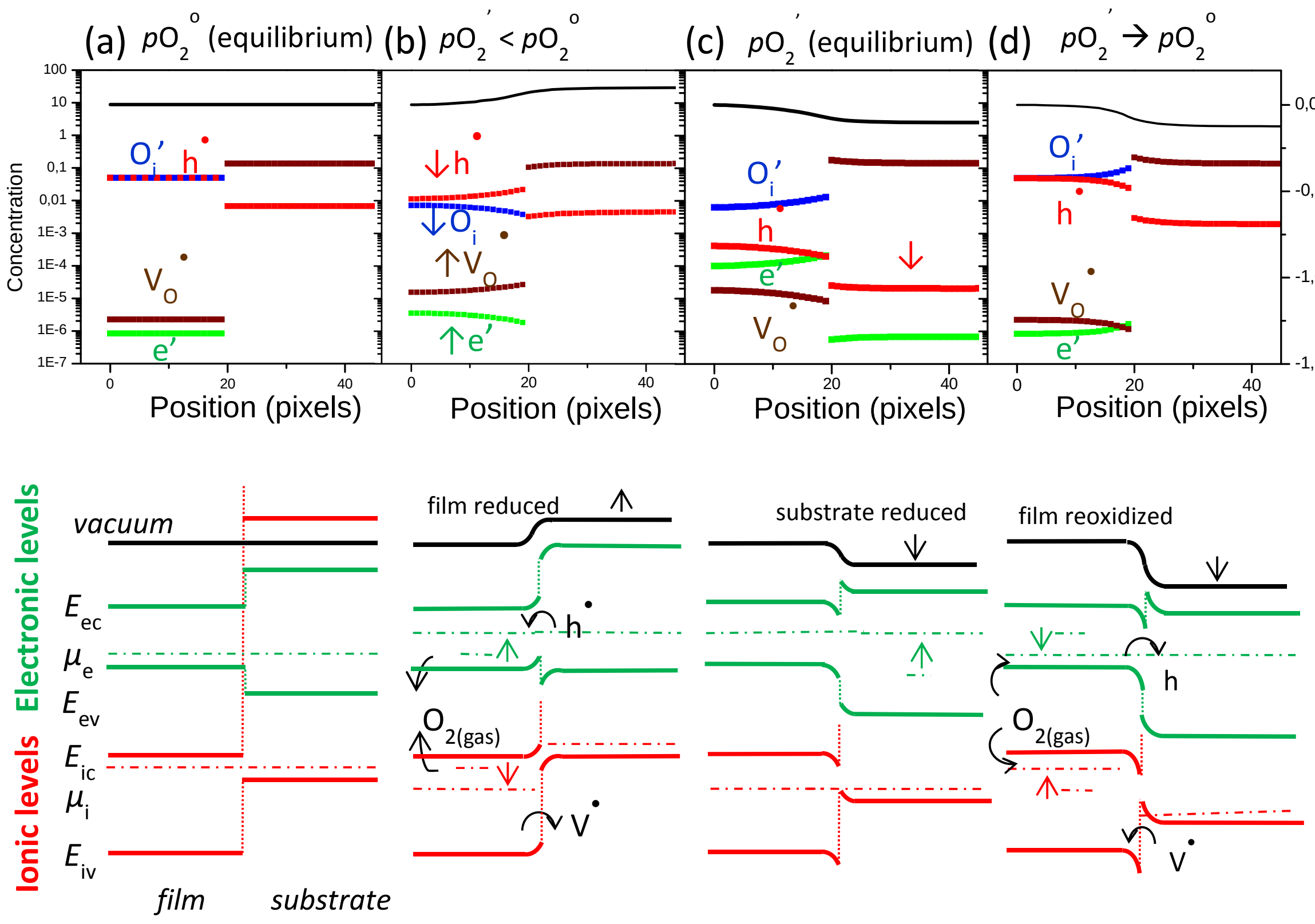

**Figure 6: (Upper panel) Defect concentration of Interstitials (blue), Electrons (green), oxygen vacancies (brown) and Holes (red) versus position obtained from simulations. x from 0 to 20 corresponds to the film and from 21 to 50 corresponds to the substrate. (Lower Panel) Corresponding band diagram for each situation (see text). In (a) starting initial conditions at oxidizing atmosphere. In (b) snapshot taken after atmosphere is switched to reducing conditions and corresponding band diagram. In (c) under reducing atmosphere after long stabilization time. In (d) in oxidation conditions after long stabilization time.**

In Fig. 6 we describe four different steps of the simulation. On the upper panel is shown the concentration of each defect and vacuum level (electrons-green, holes-red, vacancies-brown, interstitials-blue and vacuum level-black) and on the lower panel it is shown the corresponding band diagram proposed for each situation. The position of the ionic bands has been lowered to make the drawings clear enough. In (a) it is shown the initial concentration of each defect. As we initiated the system in a flat band situation, no transient phenomena is observed. In the corresponding band diagram we can observe that both, ionic and electronic chemical potentials (dashed lines) are constant across the junction, the vacuum level is constant too. In (b) it is shown the situation some iterations after the gas pressure has been reduced to $\mu_X$= -0.2. As it can be seen in the concentration profile, the film shows a tendency to desorb oxygen by reducing the amount of holes and interstitials, while the concentration in the substrate has not been affected yet. Once a chemical gradient is imposed through the junction, due to the difference in leakage currents between electronic holes and ionic vacancies, a voltage appears. Holes are accumulated at the film interface due to large injection charging positively the film. Simultaneously there is depletion of vacancies in the substrate close to the interface that stabilizes a built in voltage that oppose to the electronic current. In the steady state no net charge current is allowed, only a pure oxygen (neutral) current, so the electronic current must be equal in magnitude and opposite in direction to the ionic current established through the junction. This will be small as it is limited through injection of minority vacancies from the film into the substrate.

Using the band diagram (lower panel) we reduce the partial pressure in the film by increasing the distance between the electronic and ionic chemical potentials (by lowering $\mu_x$), while, in the other hand, the substrate is still left in an oxidized state. It is not possible to equilibrate simultaneously both chemical potentials across the interface. Holes from the substrate diffuse into the film very easily because it's large leakage current and tend to align the electronic chemical potential, while the leakage current for ionic species is expected to be much lower, so we push upward the band diagram of the substrate until we align the electronic chemical potentials. We suppose that electrons are always in equilibrium and the total ionic current is governed by the accumulation of minority vacancies in the film using eq. 7.

Once this situation is achieved, for each vacancy injected in the substrate one hole is removed. After long time, situation (c), the amount of holes in the substrate is reduced, and the built-in voltage drift in order to maintain electronic holes in equilibrium across the junction until finally stabilizes. The amount of ionic vacancies injected in the substrate is still negligible in front of the total concentration, stabilized by the doping content. The corresponding band diagram is constructed in the same way as explained before. Once the substrate is reduced we increase the distance between electronic and ionic chemical potentials. As oxygen vacancies are majority their concentration is fixed and so it is the distance between the ionic chemical potential and Ionic valence band position. On the contrary, as electrons are secondary carriers, their chemical potential can be varied, so we separate now the electronic chemical potential to achieve the new oxygen pressure. We align the electronic chemical potential but also the ionic one aligns, achieving a true equilibrium situation. It is no longer a flat band diagram. Finally, in (d) we show the concentration profile long after the oxygen pressure is increased to recover the initial one. It is observed that concentrations for each species tend to go back to their original values but there is still depletion of holes and accumulation of interstitials at the film interface. This means that there is a transient voltage across the junction, although this time the drift can be set arbitrarily long by increasing band offset or decreasing leakage current for vacancies. Experimentally it may seem that the system has reached an equilibrium situation, but it is not. Holes in the substrate increase but are still lower in concentration with respect to the initial flat band situation. Using the band diagram, we increase oxygen pressure in the film by reducing the distance between electronic and ionic chemical potentials, while the substrate is kept in a reduced state. We align the electronic chemical potential by shifting the band structure of the substrate downwards. In this situation, oxygen vacancies in the substrate need to overcome a large band offset and therefore are efficiently blocked.

This last band diagram fails in order to reproduce the electron hole concentration inside the substrate. This is because the junction cannot support a large voltage or large accumulation of vacancies near the interface, since they modify the concentration in eq. 7 and consequently the leakage current increases.

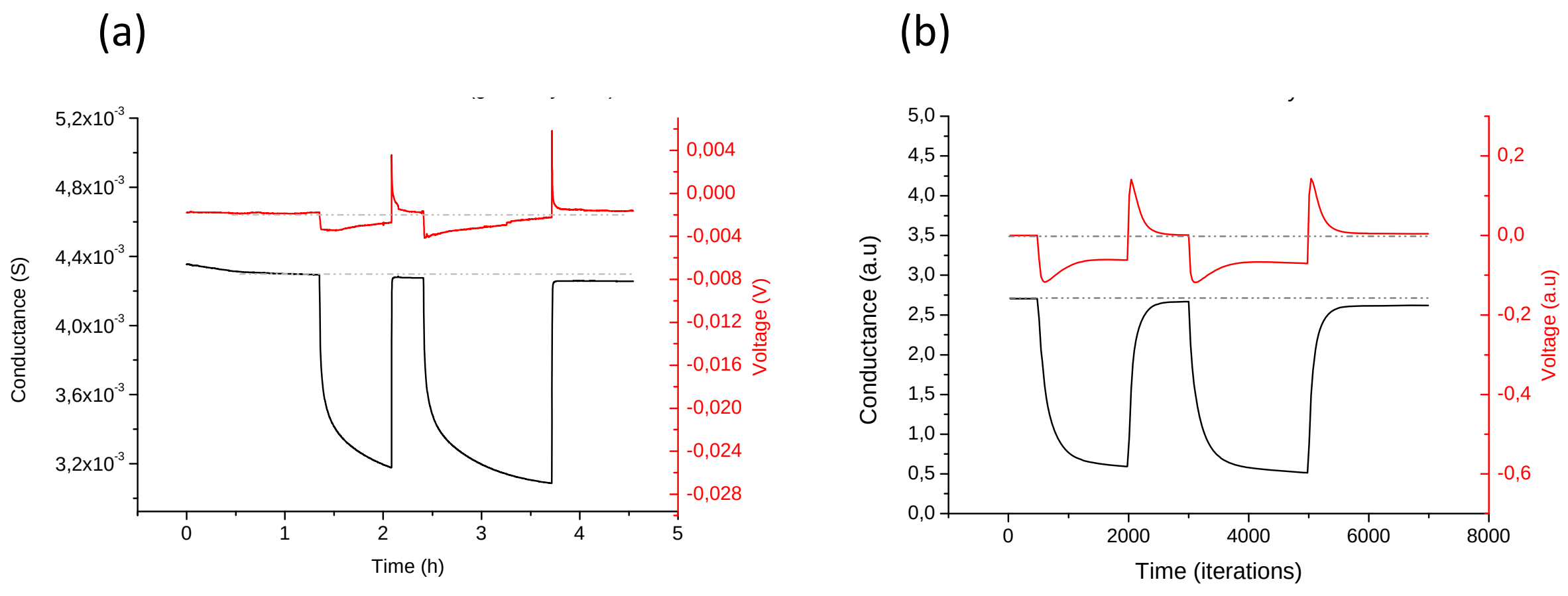

**Figure 7: (a) Experimental values of sheet conductance (black curve) and measured voltage between film and substrate (red curve) obtained on a PLD grown $La_2NiO_4$ film of around 200 nm thickness measured at 600 °C. (b) Simulated curves showing the change on conductance (integrated carrier concentration) and voltage (difference in chemical potentials of electrons)**

In fig 7 we show the measured voltage developed between the film and the substrate using a bottom electrode (in the substrate back side) during an ECR experiment in a new set of $La_2NiO_4$ samples deposited by PLD of similar characteristics. Results are compared to simulations with very good agreement.

The voltage appearing between top film and bottom electrode is just the difference on electronic chemical potential $V_{12} = \mu_e^{(2)} - \mu_e^{(1)}$. Under reducing conditions a negative voltage with respect to the substrate appears and it is maintained until the system reaches an equilibrium state. When the system is exposed to oxygen again there is a transient voltage peak (positive) although it decays very fast to zero. We attribute this voltage to the finite electronic resistance of the substrate. When a net oxygen flow is established, the electronic current is opposite to the ionic current and as the substrate offers some resistance, a voltage appears. Under reducing conditions, holes flow from the substrate to the film producing a negative persistent voltage. Under oxidation conditions the voltage is not developed meaning that there is no net electronic flow from the film to the substrate or that the oxygen flow is blocked in completely agreement with simulations.

## 4. Conclusions

The mathematical equivalence between ions and electrons, the MIES model, seems a powerful tool to describe basic phenomena of MIEC materials. It is capable of reproducing defect chemistry diagrams, as well to compute charge transport and chemical diffusion. The defect chemistry on space charged layers as well as the coupling between voltage and chemical gradients emerge in a natural way. Moreover, the introduction of absolute values of formation energies for point defects seems crucial in order to compare two materials and predict the voltage at junctions as well as charge and mass transport across them. We have shown that simulations are in agreement with band diagrams and it allows us to use the standard semiconductor concepts to calculate currents.

There are two concepts appearing with the introduction of a second kind of carrier. First, the need of using two chemical potentials, the chemical potential of the gas phase and the "charged" potential. This allows to pole the junction under two different driving forces, either chemical gradients or charge concentration gradients. The second is the role of Majority and Secondary carriers. Majority carriers fixed by doping concentrations determine the properties of the junction as the voltage, capacitance and space charge layer width in equilibrium and non-equilibrium conditions. The Secondary carriers are affected by the chemical potential of the gas phase. They may cause transient effects or small electric fields that are quickly neutralized by the majority carriers by accumulation or depletion at the interface.

It is known that the presence of extended charged defects, recombination centers or surface dipoles needs also to be taken into account in order to explain realistic currents across junctions.

Although the proposed band diagram model for this system is fully revisable and far from being realistic it explains surprisingly well the complete phenomenology observed in ECR experiments on $La_2NiO_4$ thin films on $NdGaO_3$ substrates. The simulations points to a change in built in voltage due to non-equilibrium of the substrate as the main reason for the decrease of the total conductivity observed upon cycling the atmosphere. It points also that this irreversible behavior can be produced by a rectification of one of the carriers, we expect the limiting current to be injection of minority oxygen vacancies from the film because electronic holes are majority in both sides of the junction and rectifying effects are not expected.

There are two main drawbacks of the model. First, it demands that regular oxygen from the substrate cannot go directly into interstitial sites on $La_2NiO_4$. And second, the space charged layer in these films must be comparable with film thickness in order to observe the decrease in conductivity produced by a change in built in voltage. This space charged layer with is very large compared with the short Debye lengths (of a few nanometers) calculated for the high carrier concentration in this material, but it is consistent with previously reported evidences of conductivity enhancement for thinner films and reduced oxygen diffusion near the interface [13, 19] which are compatible with an space charged region of around 50nm. This anomalously large space charge region may indicate that Oxygen interstitials and electronic holes in Nickel atoms are coupled, probably by the lattice distortion, which prevents charge separation and producing Debye lengths larger than expected for a fully ionized model. This could explain also why direct injection of regular oxygen from the substrate to interstitial sites is not permitted and the only path to reduce the substrate is through injection of minority vacancies from the film side.

We could not found other mechanism to justify all the observed phenomena in a simple way. The oxygen flow established between film and substrate seems to be the reason for the large stabilization times observed in ECR experiments done in thin films.

Recent works show evidences of other non-trivial electrochemical phenomena across junctions such as the asymmetric diffusion of $^{18}O$ in $LaAlO_3/SrTiO_3$ junctions during thin film growth, [20] or the enhanced exchange rates in a similar system like $La_2CoO_4/LaCoO_3$ bilayers [21] that could be related to the electrochemical equilibrium at interfaces at high temperatures.

Finally, we would like to mention some predictions emerging from the present work. At elevated temperatures most of the oxides develop some electronic and ionic conductivity. The model thus predicts space charged layer formation during growth of the samples and subsequent cooling down process as previously discussed by Saraf *et al.* [11]. This might have tremendous implications in some physical properties of thin films depending on deposition conditions (oxygen pressure and temperature) particularly if they are insulators at room temperature when Debye length tends to be much larger than sample thickness. In this situation the Fermi level for electrons and ions will be completely pinned by the substrate history.

Contrary to the common believe, thermal annealing of thin films do not necessarily remove oxygen vacancies if they are majority defects or stabilized by a space charged layers. Moreover, if there are other mobile species, as for instance some charged impurities, they might migrate forced by electric fields at interface regions in order to achieve flat band conditions during high temperature sintering. Thus, we predict segregation of impurities from one material to other depending of doping concentration, band alignments and oxygen pressure.

These band diagrams can be applied to explore new devices based on engineering MIEC heterojunctions. The most important functionality would be the ability of rectify neutral chemical currents by rectifying one or both of the charged carriers at interfaces. Photochemical devices or thermochemical devices can be envisioned using these band diagrams.

**Acknowledgements**

The authors would like to acknowledge the funding of this work through projects: MAT2011-29081 and CONSOLIDER-INGENIO grant CSD 2008-00023 from Spanish Ministry.

One of the authors, J.R. would like to thank Miri Markovich and Prof. Avner Rothschild from Technion University for fruitful discussions.